\documentclass[prd,twocolumn,superscriptaddress]{revtex4}
\usepackage{graphicx}
\usepackage{amsmath}
\usepackage{adjustbox}
\usepackage{gensymb}
\usepackage{amsfonts}
\usepackage{amssymb}
\usepackage{color}
\usepackage{amscd}
\usepackage{bm}
\usepackage{adjustbox}
\makeatletter
\renewcommand{\maketag@@@}[1]{\hbox{\m@th\normalsize\normalfont#1}}%
\makeatletter
\begin{document}
\title{A universal relation among Euclidean integrals for black holes in higher-derivative gravity theories}
\author{Yong Xiao}
\email{xiaoyong@hbu.edu.cn}
\affiliation{Key Laboratory of High-precision Computation and Application of Quantum Field Theory of Hebei Province,
College of Physical Science and Technology, Hebei University, Baoding 071002, China}
\affiliation{Hebei Research Center of the Basic Discipline for Computational Physics, Baoding, 071002, China}
\affiliation{Higgs Centre for Theoretical Physics, School of Mathematics, University of Edinburgh, Edinburgh, EH9 3FD, United Kingdom}
\author{Qiang Wang}
\author{Aonan Zhang}
\affiliation{Key Laboratory of High-precision Computation and Application of Quantum Field Theory of Hebei Province,
College of Physical Science and Technology, Hebei University, Baoding 071002, China}
\affiliation{Hebei Research Center of the Basic Discipline for Computational Physics, Baoding, 071002, China}
\begin{abstract}
In this paper, we establish a universal equality governing Euclidean integrals of gravitational actions in higher-derivative theories. This relation is shown to hold universally for asymptotically flat black holes in pure gravity, and is generalized to asymptotically anti-de Sitter (AdS) spacetimes through appropriate regularization. We further examine its validity in systems with matter-gravity coupling, identifying that violations occur only when matter fields exhibit pathological behaviors: divergence at the horizon or non-decaying profiles at infinity. These findings reveal fundamental constraints on gravitational thermodynamics and provide diagnostic tools for identifying ill-behaved matter configurations.
\end{abstract}

 \maketitle
 
\section{Introduction}
In the effective field theory framework for gravity, the gravitational action extends beyond the Einstein-Hilbert term to include higher-derivative terms \cite{Clifton:2011jh,Boulware:1985wk,Cardoso:2018ptl}. In $D=d+1$ ($d\geq3$) spacetime dimensions, the Lagrangian can be expressed as:
\begin{align}
  L_{\text{tot}}=\sum_{m}\alpha_m L_m(g^{\mu\nu},\nabla_\rho,R^{\mu}_{\ \nu\rho\sigma}), \label{action1}
   \end{align}
where $L_m$ denotes curvature invariants constructed from the metric $g^{\mu\nu}$, covariant derivative $\nabla_\rho$, and Riemann tensor $R^{\mu}_{\ \nu\rho\sigma}$. The coupling parameters $\alpha_m$ quantify the strength of each term, with $m$ counting the number of Riemann tensors plus half the number of explicit covariant derivatives in $L_m$. Specifically, $\alpha_1 L_1=\frac{1}{16\pi } R$ corresponds to the Einstein-Hilbert term, while $\alpha_0 L_0=-\frac{\Lambda}{8\pi}$ is the cosmological term. Thus, $\alpha_0 L_0+\alpha_1 L_1$ represents the lowest-order part of the Lagrangian. All the terms in eq.\eqref{action1} can be categorized and ordered according to the index $m\geq 0$. 

For a given action, varying it with respect to the metric $g_{\mu\nu}$ yields gravitational field equations. One can then solve these equations for black hole solutions and study their thermodynamic properties. A key tool in studying these properties is the on-shell Euclidean integral (via Wick rotation), which connects to the system's partition function and free energy \cite{hawkingpage,Dutta:2006vs,Boos:2023pyr}.

In our prior study \cite{Xiao:2022auy}, an intriguing relation among Euclidean integrals was discovered for asymptotically flat black holes ($\Lambda=0$):
\begin{align}
    \sum_{m} (2m-D)\alpha_m I_m=0, \label{xiaoformula}
\end{align}
where $I_m \equiv \int_\mathcal{M} L_m\equiv \int_{r_h}^\infty d^D x \sqrt{|g|} L_m$ denotes an integration from the horizon $r_h$ to infinity. In specific models, the action typically contains only a finite number of $\alpha_m L_m$ terms, with other coupling parameters set to 0. Consequently, the summation in eq.\eqref{xiaoformula} only includes the terms present in the action.

To illustrate formula \eqref{xiaoformula}, consider a special case in $D=4$ with
\begin{align}
\begin{split}
     & L_{\text{tot}}=\alpha_1 L_1+\alpha_2 L_2\\
      & \quad = \frac{R}{16\pi}+ \alpha_2 R_{\mu\nu\rho\sigma}R^{\mu\nu\rho\sigma},
\end{split} \label{l2ex}
\end{align} 
where $\alpha_m=0$ for $m\neq1,2$. Here eq.\eqref{xiaoformula} produces $(2\cdot 1-4)\alpha_1 I_1 + (2\cdot 2-4)\alpha_2 I_2=0$, implying $I_1\sim \int_{\mathcal{M}}R=0$. This is trivially satisfied since the Schwarzschild metric remains an exact solution of this model, which has the property $R=0$.

A non-trivial example occurs in $D=6$ with
\begin{align}
\begin{split}
   & L_{\text{tot}}=\alpha_1 L_1+\alpha_3L_3
    \\ & \quad= \frac{R}{16\pi G}+ \alpha_3 R_{\mu\nu\rho\sigma}R^{\rho\sigma\alpha\beta}R_{\alpha\beta}^{\ \ \mu\nu},
    \end{split}\label{6dl3}
\end{align}
where the higher-curvature term $\alpha_3 L_3$ modifies the Schwarzschild metric, and it leads to non-vanishing $R=-\frac{\alpha_3 28800 \pi}{r^{15}}(10m^3-7m^2r^3)$ for the perturbed black hole at $\mathcal{O}(\alpha_3)$. Yet eq.\eqref{xiaoformula} stipulates $(2\cdot 1-6)\alpha_1 I_1 + (2\cdot 3-6) \alpha_3 I_3=0$, thus enforcing
\begin{align}
I_1\sim \int_\mathcal{M} R =0, \label{6dl3result}
\end{align}
even though $R\neq 0$ here. This can be verified through explicit calculation (see the Appendix), demonstrating that eq.\eqref{xiaoformula} possesses non-trivial predictive power.

In Ref.\cite{Xiao:2022auy}, the formula \eqref{xiaoformula} was not the main focus but rather a byproduct of the primary research, receiving only a cursory explanation. 
The purpose of the present paper is to conduct a more detailed investigation of this formula. In sec.\ref{sec2}, we prove the equality for asymptotically flat cases. In sec.\ref{sec3}, we generalize it to asymptotically AdS black holes through appropriate regularizations. In sec.\ref{sec4}, we further explore whether the formula can be extended to cases involving couplings between matter fields and gravity. We summarize our findings in sec.\ref{sec5}.  Our discussion primarily focuses on stationary black holes with spherical horizons, but the analysis can be readily extended to other topological black holes with flat or hyperbolic horizons.

\section{The asymptotically-flat cases} \label{sec2}
This section presents a proof of formula \eqref{xiaoformula} for asymptotically flat spacetimes. The proof proceeds in two critical steps. First, we establish the structure of the trace of gravitational field equations. Despite the complexity of full gravitational field equations derived from varying Lagrangian \eqref{action1}, their trace (contracted with $g^{\alpha\beta}$) exhibits a surprisingly simple form \cite{Xiao:2022auy,Oliva:2010zd}:
\begin{align}
   \sum_m \left(m - \frac{D}{2}\right)\alpha_m L_m + \nabla_\mu K^\mu = 0,  \label{traceeq}
\end{align}
where $\nabla_\mu K^\mu$ denotes the total derivative term arising from curvature variations.  Second, we demonstrate that $\int_\mathcal{M}\nabla_\mu K^\mu=0$ in doing the Euclidean integrals.

As a preliminary exercise, consider the simplest Einstein-Hilbert term $\sqrt{-g}\,\alpha_1 L_1$ with $\alpha_1 =\frac{1}{16\pi}$ and $L_1=R$.  The field equation contribution $E_{\alpha\beta}$ can be extracted from the variation with respect to the metric: $\delta(\sqrt{|g|}L) \sim \sqrt{|g|}E_{\alpha\beta}\delta g^{\alpha\beta}$. For convenience, we list key variation formulas:
\begin{align}
& \delta \sqrt{-g} = \frac{1}{2}\sqrt{-g}g^{\alpha\beta}\delta g_{\alpha\beta}, \label{var1} \\
& \delta g_{\alpha\beta} = -g_{\alpha\mu}g_{\beta\nu}\delta g^{\mu\nu}, \label{var2} \\
& \delta \Gamma^{\lambda}_{\mu\nu} \! =\!  \frac{1}{2}g^{\lambda\alpha}\left(\nabla_\mu \delta g_{\alpha\nu} \!+\! \nabla_\nu \delta g_{\alpha\mu}  \!-\! \nabla_\alpha \delta g_{\mu\nu}\right) \! \sim  \!\nabla(\delta g), \label{var3} \\
& \delta R^{\lambda}_{\mu\rho\sigma} = \nabla_\rho(\delta \Gamma^{\lambda}_{\mu\sigma}) - \nabla_\sigma(\delta \Gamma^{\lambda}_{\mu\rho}) \sim \nabla\nabla(\delta g), \label{var4} \\
& \delta R_{\mu\sigma} = \delta R^{\lambda}_{\mu\lambda\sigma} \sim \nabla\nabla(\delta g). \label{var5}
\end{align}

The variation of $\sqrt{-g} R$ involves three parts: $(\delta \sqrt{-g}) R$, $\sqrt{-g}\,(\delta g^{\alpha \beta}) R_{\alpha \beta}$, and $\sqrt{-g}\, g^{\alpha \beta} \, ( \delta R_{\alpha \beta})$. Using the above formulas, the first part gives $-\frac{1}{2}\sqrt{-g}g_{\alpha\beta}R\, \delta g^{\alpha \beta}$, contributing $-\frac{D}{2}\alpha_1 L_1$ to the trace equation, after multiplying by $\alpha_1$ and contracting with $g^{\alpha\beta}$. Similarly, the second part $\sqrt{-g}R_{\alpha \beta}\, \delta g^{\alpha \beta}$ contributes $\alpha_1 L_1$ to the trace equation. The third part, involving $g^{\alpha\beta}\delta R_{\alpha \beta} \sim g \nabla \nabla \delta g $, can be simplified via two integration-by-parts steps, which can be sketched as: $g (\nabla \nabla \delta g)\sim (\nabla\nabla g)\delta g$.

Combining these, the total contribution of $\alpha_1 L_1$ to the trace equation is:
\begin{align}
    \left(1 - \frac{D}{2}\right)\alpha_1 L_1 + \nabla^\mu K_\mu, \label{traceL1}
\end{align}
The last term is a total derivative term originating from $\delta R_{\alpha\beta}$. For the simplest case $\alpha_1 L_1=\frac{1}{16\pi}R$, it simply vanishes because $ \nabla^\mu K_{\mu}\sim \nabla^{\mu}\nabla^{\nu }g_{\mu\nu}=0$.

In fact, higher-derivative terms in the Lagrangian also lead to a pattern similar to that in eq.\eqref{traceL1}. To illustrate this, consider a term $\alpha_5 L_5 = \alpha_5 R R_{\mu\nu} R^{\mu\eta\rho\sigma} \nabla_\lambda \nabla^\lambda R^\nu_{\ \eta\rho\sigma}$ and examine its contribution to the trace equation. The variation of $\sqrt{|g|}\alpha_5 L_5$ still consists of three parts. The first part comes from $(\delta\sqrt{|g|})\alpha_5 L_5$, and its contribution to the trace equation is $-\frac{D}{2} \cdot \alpha_5 L_5$. The second part arises from the variation of $g^{\alpha\beta}$ while keeping $R^\lambda_{\ \mu\rho\sigma}$ fixed. We express $\alpha_5 L_5$ in the following form:
\begin{align}
\begin{split}
& \alpha_5 L_5 = \alpha_5 \, g^{\mu_1\mu_2} g^{\mu_3\eta} g^{\mu_4\rho} g^{\mu_5\sigma} g^{\mu_6\mu_7} \\
& \quad \cdot \, R_{\mu_1\mu_2} R_{\mu\nu} R^\mu_{\ \mu_3\mu_4\mu_5} \nabla_{\mu_6} \nabla_{\mu_7} R^\nu_{\ \eta\rho\sigma}.
\end{split} \label{l5ex}
\end{align}
Obviously, eq.\eqref{l5ex} contains five factors of $g^{\alpha\beta}$. Varying each of them introduces five different terms into the gravitational field equations. Fortunately, after contracting these five terms with the metric, they revert to the form of $\alpha_5 L_5$. Thus, the contribution of the second part to the trace equation is $5 \cdot \alpha_5 L_5$. The third part comes from the variation of curvature terms and covariant derivative operators, which transforms into a total derivative through integration by parts. Specifically, the derivation for the variation of curvature terms can be sketched as: $A \delta R \sim A (\nabla\nabla \delta g) \sim (\nabla\nabla A) \delta g$. The variation of covariant derivatives can be sketched as: $A (\delta\nabla) B \sim A (\delta\Gamma) B \sim AB (\nabla\delta g) \sim (\nabla AB) \delta g$. For technical details on handling these variations, refer to Ref.\cite{Biswas:2013cha}. Finally, summing up the contributions of these three parts, the total contribution to the trace equation is:
\begin{align}
\left(5 - \frac{D}{2}\right) \alpha_5 L_5 + \nabla_\mu K^\mu.
\end{align}
Here, $K_\mu$ has a relatively complex form but is nonetheless an expression composed solely of curvature tensors and covariant derivatives.

Generalizing, for any $\alpha_m L_m$, its trace equation contribution has three parts: $-\frac{D}{2}\cdot \alpha_m L_m$ (from $\delta\sqrt{|g|}$), $m\cdot\alpha_m L_m$ (from varying $g^{\alpha\beta}$), and $\nabla_\mu K^\mu$ (from varying curvatures or covariant derivatives). Thus, the trace equation takes the form \eqref{traceeq}.

It is worth noting that in Ref.\cite{Oliva:2010zd}, the trace equation \eqref{traceeq} was derived via an independent method using scaling analysis. Moreover, Ref.\cite{Oliva:2010zd} concretely analyzed all possible $\alpha_3 L_3$ terms, which include 12 linearly independent terms such as $R^{abcd} R_{cdef} R^{ef}_{ab}$, $R^{abcd} R_{ac} R_{bd}$, and $\nabla_\lambda R_{abcd} \nabla^\lambda R^{abcd}$, etc. For the specific theories, they identified combinations of coupling coefficients that exactly cancel the total derivative term, in order to simplify the trace equation to the form
\begin{align}
\sum_m \left(m - \frac{D}{2}\right) \alpha_m L_m = 0. \label{traceeq0}
\end{align}
Such theories possess favorable properties, allowing for exact black hole solutions even when these curvature combinations do not belong to the Lovelock class of theories. This is certainly an interesting class of theories, valuable in its own right. However, in this paper, we proceed with a general analysis without focusing specifically on this type of theory.

Next, to complete the proof, we show $\int_\mathcal{M} \nabla_\mu K^\mu = 0$ for asymptotically-flat black holes, even if $\nabla_{\mu}K^\mu \neq 0$ in general. By Gauss's law, for a manifold $\mathcal{M}$ bounded by $S_\infty$ (infinity) and $S_h$ (horizon):
\begin{align}
    \int_\mathcal{M} \nabla_\mu K^\mu = \int_{S_\infty} n_\mu K^\mu - \int_{S_h} n_\mu K^\mu, \label{kflat}
\end{align}
where $n^\mu$ is the radial unit normal. Since $K^\mu$ is an expression composed of curvature terms and covariant derivatives, and for asymptotically flat spacetimes the curvature terms vanish at infinity, so there is $\int_{S_\infty} n_\mu K^\mu = 0$. On the other hand, the event horizon is located at the tip of a cigar-like shape of the Euclidean manifold, so the flux $\int_{S_h} n_\mu K^\mu = 0$. From the mathematical perspective, this is a natural result caused by $\int_{S_{r_h}} g^{rr} n_r K_r = 0$ with $g^{rr}(r)\big|_{r=r_h} = 0$, provided that the spacetime curvature contained in $K^\mu$ has a regular behavior at the horizon.

Therefore, by performing the Euclidean integral on both sides of formula \eqref{traceeq} and multiplying by a factor of 2, we obtain:
\begin{align}
    \sum_m (2m - D)\,\alpha_m \int_{\mathcal{M}} L_m = 0
\end{align}
Thus, we have completed the proof of formula \eqref{xiaoformula} for asymptotically flat black holes.

\section{The asymptotically-AdS cases} \label{sec3}
\subsection{General analysis}

In this section, we extend the equality relation \eqref{xiaoformula} to asymptotically AdS spacetimes, where the gravitational action includes the cosmological term $\alpha_0 L_0$ with $\alpha_0 = -\frac{\Lambda}{8\pi}$ and $L_0 = 1$.

Varying the cosmological term $\sqrt{|g|}\alpha_0 L_0$ contributes $\left(0 - \frac{D}{2}\right)\alpha_0 L_0$ to the trace equation. Thus, the trace equation retains the same structure as in asymptotically flat spacetimes:
\begin{align}
   \sum_m \left(m - \frac{D}{2}\right)\alpha_m L_m + \nabla_\mu K^\mu = 0.  \label{traceeqads}
\end{align}

We now analyze the Euclidean integrals $\int_\mathcal{M} L_m$. A key subtlety in AdS spacetimes is that direct computation of these integrals yields divergences. To handle this in black hole thermodynamics, one needs to regularize the Euclidean integrals using a subtraction method, which removes contributions from the pure AdS background \cite{hawkingpage,Dutta:2006vs}. The regularized Euclidean integral is defined as
\begin{align}
  I_m \equiv \int_{\mathcal{M}}^{(\text{Reg.})} L_m,
\end{align}
where the regularized integral is given by
\begin{align}
  \int_{\mathcal{M}}^{(\text{Reg.})} \equiv \beta \lim_{r_c \to \infty} \left( \int_{\mathcal{V}_{r_c}}^{(\text{BH})} - k_c \cdot \int_{\mathcal{V}_{r_c}}^{(\text{AdS})} \right). \label{regIntegral}
\end{align}
In this definition, $\beta \equiv 1/T$ denotes the periodicity of Euclidean time, $r_c$ is a radial cutoff at large distances, and $\mathcal{V}_{r_c}$ represents the volume up to $r_c$. 

In eq.\eqref{regIntegral}, the first integral is evaluated over the black hole spacetime (from the horizon $r = r_h$ to $r_c$). The second integral corresponds to the pure AdS background (from $r = 0$ to $r_c$). When subtracting the two, their divergent behaviors at $r_c\rightarrow \infty$ cancel each other, thus yielding a finite result. By the way, the factor $k_c$ is chosen according to the requirements of the problem under study. Practical choices for $k_c$ include $k_c = 1$ and $k_c = \sqrt{\frac{g_{tt}^{\text{BH}}(r_c)}{g_{tt}^{\text{AdS}}(r_c)}}$.
 For the former choice, $\int_{\mathcal{M}}^{(\text{Reg.})} L$ can be regarded as a straightforward generalization of the Euclidean integral of the asymptotically-flat case, and the integration result is related to the Komar mass and Smarr relation \cite{Xiao:2023lap}. For the latter choice, $\int_{\mathcal{M}}^{(\text{Reg.})} L$ directly gives the partition function of the system, meaning that the calculation result automatically includes contributions from boundary terms such as the Gibbons-Hawking-York term \cite{hawkingpage,Dutta:2006vs}. As will be seen in the subsequent general analysis, our conclusions are independent of the choice of $k_c$.
 
We propose the following generalization of eq.\eqref{xiaoformula} to asymptotically AdS black holes:
\begin{align}
  \sum_m (2m - D) \alpha_m I_m = 0, \label{xiaoforads}
\end{align}
where $I_m \equiv \int_{\mathcal{M}}^{(\text{Reg.})} L_m$. To illustrate this, we integrate the trace equation \eqref{traceeqads} and show that $\int_{\mathcal{M}}^{(\text{Reg.})} \nabla_\mu K^\mu = 0$. Using Gauss's theorem, we have
\begin{align}
  \int_{\mathcal{M}}^{(\text{Reg.})} \nabla_\mu K^\mu &= \left( \int_{S_\infty}^{(\text{BH})} n_\mu K^\mu - \int_{S_h}^{(\text{BH})} n_\mu K^\mu \right) \notag \\
  &\quad - k_c \cdot \int_{S_\infty}^{(\text{AdS})} n_\mu K^\mu. \label{regk}
\end{align}
In the case of pure gravity, as $K^{\mu}$ contains a free index, its construction must include derivatives of Riemann tensor. It is known that the covariant derivative of Riemann tensor vanishes identically for pure AdS spacetime. Therefore, the third term in eq.\eqref{regk} is directly zero, regardless of the choices of $k_c$. Similarly, because $K^\mu$ involves derivatives of Riemann tensor, the first term also decays rapidly to zero as $r_c \to \infty$. For example, in the simplest case where $K^\mu = \nabla_\nu R^{\mu\nu}$, practical evaluation shows that its decay behavior is equal to or faster than $\frac{\alpha_m}{r^{2(m-1)+(D-3)}}$ ($m \geq 2$). This is physically understandable: since the influence of the mass source vanishes at infinity, the contribution of this term should naturally approach that of the pure AdS background. Finally, for the second term, its value at the horizon must also be zero, for the same reason as analyzed below eq.\eqref{kflat}. In total, there is  $\int_{\mathcal{M}}^{(\text{Reg.})} \nabla_\mu K^\mu = 0$.

\subsection{Concrete example}
To exemplify the validity of the generalized equality \eqref{xiaoforads} in asymptotically AdS cases, we consider a model described by the Lagrangian
\begin{align}
\begin{split}
L_{\text{tot}} &= \frac{1}{16\pi}\left(R - 2\Lambda\right) + \alpha_2 R_{\mu\nu\rho \sigma}R^{\mu\nu \rho \sigma} \\
&\quad + \alpha_3 R_{\mu\nu}^{\ \ \rho\sigma}R_{\rho\sigma}^{\ \ \alpha\beta}R_{\alpha\beta}^{\ \ \mu\nu},
\end{split} \label{lagrangian23}
\end{align}
where $\Lambda = -\frac{3}{l^2}$ and $l$ denotes the AdS length scale. The terms can be classified as follows: 
\begin{align}
   & \alpha_0 L_0 = -\frac{\Lambda}{8\pi},\\
    & \alpha_1 L_1 = \frac{R}{16\pi},\\
    & \alpha_2 L_2 = \alpha_2 R_{\mu\nu}R^{\mu\nu},\\
    &\alpha_3 L_3 = \alpha_3 R_{\mu\nu}^{\ \ \rho\sigma}R_{\rho\sigma}^{\ \ \alpha\beta}R_{\alpha\beta}^{\ \ \mu\nu},
\end{align}
which are respectively the cosmological term, the Einstein-Hilbert term, and the quadratic and cubic curvature terms.

In $D=4$ dimensions, the black hole solution can be derived perturbatively around the AdS-Schwarzschild metric, up to first order in the couplings $\alpha_2$ and $\alpha_3$. The solution is given by \cite{Xiao:2023lap}
\begin{align}
ds^2 = -f(r)dt^2 + \frac{1}{g(r)}dr^2 + r^2 d\Omega_2, \label{metricFG}
\end{align}
where $d\Omega_2$ is the line element of the 2-sphere, and the functions $f(r)$ and $g(r)$ take the form $f(r) = 1 - \frac{2 M}{r} + \frac{r^2}{l^2}+f^{(1)}(r)$, and $g(r) = 1 - \frac{2 M}{r} + \frac{r^2}{l^2}+g^{(1)}(r)$, where
\begin{align}
\begin{split}
f^{(1)}(r)  \!= \! \frac{64 \pi \alpha_3 }{l^6 r^7} \left(10 l^6 M^3 \!+ \! 12 l^4 M^2 r^3  \!- \! 6 l^2 M r^6  \!+ \! r^9\right),
\end{split}
\end{align}
and
\begin{align}
\begin{split}
g^{(1)}(r) &=  \frac{64 \pi \alpha_3 }{l^6 r^7} \left(54 l^6 M^2 r - 98 l^6 M^3 + 66 l^4 M^2 r^3 \right. \\
&\quad \left. - 6 l^2 M r^6 + r^9\right).
\end{split}
\end{align}
The relation between the mass parameter $M$ and the horizon radius $r_h$ is obtained from the condition $f(r_h) = g(r_h) = 0$, yielding $M = \frac{r_h}{2} + \frac{r_h^3}{2 l^2}  + \frac{8 \pi \alpha_3 }{l^6 r_h^3}\left(5 l^6 + 27 l^4 r_h^2 + 27 l^2 r_h^4 + 9 r_h^6\right)$.

Using this metric, we can explicitly calculate the regularized integrals with $k_c = 1$, resulting in:
\begin{align}
\begin{split}
\alpha_0 I_0 = -\frac{\beta r_h^3}{2 l^2}  - \frac{216 \pi \alpha_3 \beta }{l^6 r_h} \left(l^2 + r_h^2\right)^2,
\end{split}
\end{align}
\begin{align}
\begin{split}
\alpha_1 I_1 \!=\! \frac{\beta r_h^3}{l^2} \!+\! \frac{8 \pi \alpha_3 \beta }{l^6 r_h^3}\left(l^6 \!+\! 45 l^4 r_h^2 \!+\! 87 l^2 r_h^4 \!+\! 51 r_h^6\right),
\end{split}
\end{align}
\begin{align}
\alpha_2 I_2 = \frac{16 \pi \alpha_2 \beta }{l^4 r_h}\left(l^4 + 2 l^2 r_h^2 - r_h^4\right),
\end{align}
\begin{align}
\begin{split}
\alpha_3 I_3 = \frac{8 \pi \alpha_3 \beta }{l^6 r_h^3} \left(l^6 - 9 l^4 r_h^2  - 21 l^2 r_h^4 - 3 r_h^6\right).
\end{split}
\end{align}
We can verify that they indeed satisfy the relation
\begin{align}
\begin{split}
&(2 \cdot 0 - 4)\alpha_0 I_0 + (2 \cdot 1 - 4)\alpha_1 I_1 + (2 \cdot 2 - 4)\alpha_2 I_2 \\
&\quad + (2 \cdot 3 - 4)\alpha_3 I_3 = 0,
\end{split}
\end{align}
confirming the validity of eq.\eqref{xiaoforads} in this specific example.

\section{Actions with matter fields coupling to gravity} \label{sec4}

\subsection{General analysis}

In higher-derivative theories of gravity, we often encounter scenarios where matter fields are coupled to gravity. Thus, we are interested in whether the equality relations \eqref{xiaoformula} and \eqref{xiaoforads} can be generalized to such cases.

As seen earlier, the key to proving these equalities lies in showing that $\int_{S} n_{\mu}K^{\mu}$ vanishes both at the horizon and at infinity. In contrast to the pure gravity case, once matter fields are non-minimally coupled to gravity (i.e., they couple to curvature terms), $K^{\mu}$ will include contributions from matter fields. To ensure $\int_S n_{\mu}K^{\mu}$ still vanishes, we must impose constraints on the properties of matter fields, i.e., matter fields must not diverge at the horizon, and they must decay appropriately at infinity.

Physically meaningful matter fields with well-behaved properties naturally satisfy these requirements, such as the gauge field $F_{\mu\nu}$. However, for scalar fields or other models lacking proper symmetry constraints, pathological behaviors often arise, violating our necessary conditions. We have indeed identified specific cases where such violations occur. In these situations, the breakdown of eqs.\eqref{xiaoformula} and \eqref{xiaoforads} can conversely signal anomalous behaviors of the matter fields. Specific examples and counter-examples of eqs.\eqref{xiaoformula} and \eqref{xiaoforads} are presented in the following subsections.

\subsection{Example: higher derivative Einstein-Maxwell theory}
We consider the Einstein-Maxwell theory with higher derivative interactions described by the Lagrangian \cite{Cremonini:2019wdk}
\begin{align}
\begin{split}
L &= \frac{1}{16\pi } \bigg(R - 2\Lambda - \frac{1}{4}F^2 + c_1 R_{\mu\nu\rho\sigma}R^{\mu\nu\rho\sigma} \\
&\quad + c_2 R_{\mu\nu\rho\sigma}F^{\mu\nu}F^{\rho\sigma} + c_3 (F^2)^2 + c_4 F^4 \bigg), \label{actmaxwell}
\end{split}
\end{align}
where $F^2\equiv F_{\mu\nu}F^{\mu\nu}$, $F^4\equiv F_{\mu\nu}F^{\nu}_{\ \alpha}F^{\alpha}_{\ \beta}F^{\beta \mu}$. Here $F_{\mu\nu}$ is the Maxwell field strength tensor, and $c_i$'s are coupling parameters for higher-derivative terms.

We classify the terms by the index $m$, which now counts the number of curvature tensors plus half the number of $\nabla_\mu$ and the indices carried by the matter fields. This classification of the higher-derivative terms yields
\begin{align}
   & L_2 \sim -\frac{1}{4}F^2 + c_1 R_{\mu\nu\rho\sigma}R^{\mu\nu\rho\sigma},\\
    & L_3 \sim c_2 R_{\mu\nu\rho\sigma}F^{\mu\nu}F^{\rho\sigma},\\
   & L_4 \sim c_3 (F^2)^2 + c_4 F^4.
\end{align} Correspondingly, the trace equation for this system can be written as
\begin{align}
\sum_m \left(m - \frac{D}{2}\right)\alpha_m L_m + \nabla_\mu K^{\mu} = 0.
\end{align}
As before, the total derivative term $\nabla_\mu K^\mu$ arises from variations of curvature terms. Specifically, it originates from contributions like $c_1 R_{\mu\nu\rho\sigma} (\delta R^{\mu\nu\rho\sigma})$ and $c_2 (\delta R_{\mu\nu\rho\sigma}) F^{\mu\nu}F^{\rho\sigma}$. The form of $K^{\mu}$ can be sketched as:
\begin{align}
K^{\mu} \sim c_1 \nabla R_{....} + c_2 \nabla (F_{..} F^{..}).
\end{align}

In $D=4$ dimensions, the black hole solution at first order in the couplings $c_i$'s was derived in Ref.\cite{Cremonini:2019wdk}. The metric takes the form $f(r)= 1 - \frac{m}{r} + \frac{q^2}{4 r^2} + \frac{r^2}{l^2} +f^{(1)}(r)$ and $ g(r)=1 - \frac{m}{r} + \frac{q^2}{4 r^2} + \frac{r^2}{l^2}+g^{(1)}(r)$, where
\begin{align}
\begin{split}
&f^{(1)}(r) = \frac{1}{20 l^2 r^6}\Big [\!-\!2 c_1 q^2 \left(\!-\! 10 m l^2 r \!+\! q^2 l^2 \!+\! 20 l^2 r^2 \right. \\ &\quad \left. + 40 r^4\right) - c_2 q^2 \left( 10 r l^2 (2 r - m) + q^2 l^2 + 100 r^4\right)\\
&\quad  - 8 c_3 l^2 q^4 - 4 c_4 l^2 q^4 \Big ],
\end{split}
\end{align}
and
\begin{align}
\begin{split}
&g^{(1)}(r)\!=\! \frac{1}{10 l^2 r^6}\Big[ c_1 (-6 l^2 q^4 \!+\! 30 l^2 m q^2 r \!-\! 40 l^2 q^2 r^2 \\   &\quad - 60 q^2 r^4) 
 + c_2 (-8 l^2 q^4 + 35 l^2 m q^2 r - 40 l^2 q^2 r^2 \\&\quad  - 80 q^2 r^4) 
- 4 c_3 l^2 q^4 - 2 c_4 l^2 q^4 \Big].
\end{split}
\end{align}
The relationship between the mass parameter $m$ and the horizon radius $r_h$ is given by $m= r_h + \frac{r_h^3}{l^2} + \frac{q^2}{4 r_h} +m^{(1)}$, where
\begin{align}
\begin{split}
& m^{(1)} = \frac{1}{40 l^2 r_h^5} \Big[ 6 c_1 l^2 q^4 - 40 c_1 l^2 q^2 r_h^2 - 120 c_1 q^2 r_h^4 \\
&\quad + 3 c_2 l^2 q^4 - 20 c_2 l^2 q^2 r_h^2 - 180 c_2 q^2 r_h^4 \\
&\quad - 16 c_3 l^2 q^4 - 8 c_4 l^2 q^4 \Big].
\end{split}
\end{align}
We are mainly concerned about the behavior of the matter field $F_{\mu\nu}$, whose non-zero components are given by
\begin{align}
\begin{split}
& F_{tr} =-F_{rt}= \frac{q}{r^2} + \frac{1}{2 r^6 l^2} \Big[2 c_1 l^2 q^3 
+ c_2 \left(-9 l^2 q^3  \right. \\ &\quad \left. + 16 l^2 m q r - 16 q r^4\right) - 32 c_3 l^2 q^3 - 16 c_4 l^2 q^3 \Big].
\end{split}
\end{align}
The Maxwell field satisfies our requirements: it decays rapidly at infinity and remains finite at the horizon. Concretely, we find that $K_{\mu} \sim \nabla_\rho (F_{\mu\nu}F^{\nu\rho}) \sim 1/r^5$. This ensures the validity of eq.\eqref{xiaoforads}.

For explicit verification, we calculate the regularized integrals $I_m \equiv \int^{\text{(reg.)}}_\mathcal{M} L_m$ with $k_c=1$. The results are respectively
\begin{align}
\begin{split}
\alpha_0 I_0 &= -\frac{8 \pi r_h^3 \beta}{l^2} + \frac{12 \pi q^2 \beta}{l^2 r_h}(2 c_1 + 3 c_2),
\end{split}
\end{align}
\begin{align}
\begin{split}
 \alpha_1 I_1 & = \frac{16 \pi r_h^3 \beta}{l^2} 
- \frac{\pi q^2 \beta}{5 l^2 r_h^5}\Big[240 c_1 r_h^4 + c_2 \left( 20 l^2  r_h^2 \right. \\ &\quad \left. - 7  q^2 l^2   + 300 r_h^4\right) 
 - 32 c_3 l^2 q^2 - 16 c_4 l^2 q^2 \Big],
\end{split}
\end{align}
\begin{align}
\begin{split}
\alpha_2 I_2 & \!=\! \frac{2 \pi q^2 \beta}{r_h} + \frac{\pi \beta}{5 l^4 r_h^5} \Big[ c_1 \left(l^4 \left(6 q^4 \!-\! 20 q^2 r_h^2 \!+\! 80 r_h^4\right) \right. \\
&\quad \left. - 20 l^2 r_h^4 \left(q^2 - 8 r_h^2\right) - 80 r_h^8\right) \\
&\quad + c_2 \left(-11 l^4 q^4 + 40 l^4 q^2 r_h^2 - 120 l^2 q^2 r_h^4\right) \\
&\quad - 64 c_3 l^4 q^4 - 32 c_4 l^4 q^4 \Big],
\end{split}
\end{align}
\begin{align}
\begin{split}
\alpha_3 I_3 &= \frac{c_2 \pi q^2 \beta}{5 l^2 r_h^5} \left(l^2 \left(7 q^2 - 20 r_h^2\right) + 60 r_h^4\right),
\end{split}
\end{align}
\begin{align}
\alpha_4 I_4 &= \frac{8 \pi q^4 \beta}{5 r_h^5} (2 c_3 + c_4).
\end{align}
Even for such complex calculation results, it can be verified that they still satisfy the simple equality \eqref{xiaoforads}, namely
\begin{align}
\begin{split}
&(2 \cdot 0 - 4) \alpha_0 I_0 + (2 \cdot 1 - 4) \alpha_1 I_1 + (2 \cdot 2 - 4)\alpha_2 I_2 \\
&\quad + (2 \cdot 3 - 4) \alpha_3 I_3 + (2 \cdot 4 - 4)\alpha_4 I_4 = 0.
\end{split}
\end{align}

\subsection{(Counter-)example: gravity coupled with a background Kalb-Ramond field}

Next, we seek counter-examples to the equalities \eqref{xiaoformula} and \eqref{xiaoforads}. As noted earlier, such counter-examples arise solely for unconventional matter field configurations.

In this subsection, we consider cases where matter fields fail to decay at infinity, potentially violating the equalities. Below we provide an example involving the Kalb-Ramond field, an antisymmetric tensor field whose non-vanishing vacuum expectation value leads to anomalous asymptotic behavior. 

The Lagrangian of the Kalb-Ramond field coupled to Einstein gravity is given by \cite{Altschul:2009ae}
\begin{align}
\begin{split}
L &= \frac{1}{16\pi} \big(R - 2\Lambda + \xi_1 B_{\mu\nu}B^{\mu\nu}R + \xi_2 B^{\rho \mu}B^{\nu}_{\ \mu}R_{\rho \nu}\big) \\
&\quad - \frac{1}{12}H^{\mu\nu\rho}H_{\mu\nu\rho} - V(B^{\mu\nu}B^{\mu\nu} + b^2), \label{krfield}
\end{split}
\end{align}
where $B_{\mu\nu}$ is the Kalb-Ramond field, $H_{\mu\nu \rho} \equiv \nabla_{[\mu}B_{\nu \rho]}$ denotes its field strength, and the constant $b^2$ controls the  expectation value of $B_{\mu\nu}$ to satisfy vacuum condition $V(X) = 0$, where $X=B^{\mu\nu}B^{\mu\nu} + b^2$. For simplicity, we shall set $\xi_1=0$ in the following.

We analyze a black hole solution obtained by Ref.\cite{Liu:2025fxj} in $D=4$ dimensions. For a potential $V(X) = \frac{\lambda}{2}X$ with $\lambda = \frac{\Lambda \xi_2}{8\pi \left(1 - \frac{b^2 \xi_2}{2}\right)}$, the metric takes the form \eqref{metricFG} with:
\begin{align}
f(r) = \frac{1}{1 - \frac{b^2 \xi_2}{2}} - \frac{2 m}{r} - \frac{\Lambda r^2}{3 \left(1 - \frac{b^2 \xi_2}{2}\right)},
\end{align}
and $g(r) = f(r)$. The non-zero components of the Kalb-Ramond field's vacuum expectation value are $\langle B_{10} \rangle = -\langle B_{01} \rangle = \frac{b}{\sqrt{2}}$. The relation between the mass parameter $m$ and horizon radius $r_h$ follows from $f(r_h) = 0$:
\begin{align}
m = \frac{r_h \left(3 - \Lambda r_h^2\right)}{6 - 3 b^2 \xi_2}.
\end{align}

Observing the Lagrangian and taking into account the explicit form of $V(X)$, we can classify the terms by their curvature and matter content:
\begin{align}
 &   \alpha_0 L_0 = -\frac{\Lambda}{8\pi} - \frac{1}{2}\lambda b^2,\\
   & \alpha_1 L_1 = \frac{R}{16\pi},\\
&   \alpha_2 L_2 = -\frac{1}{2}\lambda B^{\mu \nu} B^{\mu\nu},\\
& \alpha_3 L_3 = \frac{1}{16\pi}\xi_2 B^{\rho \mu}B^{\nu}_{\ \mu}R_{\rho\nu} - \frac{1}{12}H^{\mu\nu\rho}H_{\mu\nu\rho}.
\end{align}
Then the trace equation can be easily written out as
\begin{align}
\sum_m \left(m - \frac{D}{2}\right) \alpha_m L_m + \nabla_\mu K^{\mu} = 0,
\end{align}
where $\nabla_\mu K^{\mu}$ arises from variations of curvature terms coupled to the Kalb-Ramond field, which can be sketched as $B^{\rho\mu}B^{\nu}_{\ \mu}\delta R_{\rho \nu} \sim BB (\nabla\nabla \delta g) \sim \nabla \nabla (BB) \cdot \delta g$. Direct calculation shows $K^{\mu}$ includes contributions from $\nabla^{\nu}(B^{\mu\rho}B_{\nu\rho})$, leading to:
\begin{align}
\begin{split}
\int_{S_r} n_{\mu} K^{\mu} &\sim \frac{b^2 \xi_2 \left(3 m \left(b^2 \xi_2 - 2\right) - \Lambda r^3 + 3 r\right)}{6 - 3 b^2 \xi_2}.
\end{split} 
\end{align}
At the horizon $r = r_h$, this integral vanishes and thus poses no issue. However, the integral diverges as $r \to \infty$, which is related to the fact that the Kalb-Ramond field does not decay at $r \to \infty$ (it has a non-zero component $\langle B_{10} \rangle = \frac{b}{\sqrt{2}}$ as mentioned above). Even after regularization by subtracting the background contributions (with $m=0$), the result remains non-zero:
\begin{align}
\begin{split}
    \int^{(reg.)}_{S_\infty} n_{\mu} K^{\mu} & \!= \! \lim_{r_c \to \infty} \big( \int^{(BH)}_{S_{r_c}} n_{\mu} K^{\mu} \!-\! \int^{(AdS)}_{S_{r_c}} n_{\mu} K^{\mu} \big) \\
   &  \sim b^2\xi_2 m.
\end{split}\label{krnk}
\end{align}
The result is independent of $\Lambda$, so it implies that this term remains non-zero even in the asymptotically-flat case. Consequently, for this black hole solution, the formulas \eqref{xiaoformula} and \eqref{xiaoforads} fail to hold, whether in asymptotically flat or asymptotically AdS spacetimes.

For concrete verification, substituting the metric to calculate the regularized Euclidean integral $I_m$ with $k_c=1$ yields
\begin{align}
\alpha_0 I_0 &= \frac{\Lambda r_h^3 \beta}{6 - 3 b^2 \xi_2},
\end{align}
\begin{align}
\alpha_1 I_1 &= \frac{\left(3 b^2 r_h \xi_2 - 4 \Lambda r_h^3\right)\beta}{12 - 6 b^2 \xi_2},
\end{align}
\begin{align}
\alpha_2 I_2 &= \frac{b^2 \Lambda r_h^3 \xi_2 \beta}{6 \left(b^2 \xi_2 - 2\right)},
\end{align}
\begin{align}
\alpha_3 I_3 &= \frac{b^2 \Lambda r_h^3 \xi_2 \beta}{12 - 6 b^2 \xi_2}.
\end{align}
Then we indeed find
\begin{align}
\sum_m (2m - D) \alpha_m I_m = -\beta\, b^2 \xi_2 m \neq 0,
\end{align}
thus confirming the failure of the equality for theories with non-decaying Kalb-Ramond fields. This breakdown highlights the importance of proper asymptotic behavior for matter fields in preserving gravitational thermodynamics relations.

\subsection{(Counter-)example: black hole with scalar hair}

We further search for examples where matter fields diverge at the horizon, potentially violating eqs.\eqref{xiaoformula} or \eqref{xiaoforads}.

Consider the Lagrangian \cite{Herdeiro:2015waa}
\begin{align}
L_{\text{tot}} = \frac{1}{4\pi } \left( \frac{R}{4} - \frac{1}{2}\nabla_\mu\Phi \nabla^{\mu}\Phi - \frac{1}{12}R\Phi^2 \right). \label{actscalar}
\end{align}
All terms in this Lagrangian can be classified as $\alpha_1 L_1$. So the trace equation takes the form
\begin{align}
\left(1 - \frac{D}{2}\right)L_{\text{tot}} + \nabla_\mu K^{\mu} = 0.
\end{align}
The total derivative term $\nabla_\mu K^{\mu}$ arises only from variations of terms coupled to the Ricci scalar $R$, specifically from $-\frac{1}{12}(\delta R)\Phi^2$. This can be sketched as $(\nabla\nabla\delta g)\Phi^2 \sim (\nabla\nabla\,\Phi^2)\delta g$, implying $K^{\mu} \sim \nabla^\mu \Phi^2$.

In $D=4$ dimensions, the theory has the black hole solution with the form \cite{bekensteinScalar,Herdeiro:2015waa}
\begin{align}
ds^2 = -f(r)dt^2 + \frac{1}{g(r)}dr^2 + r^2 d\Omega_2,
\end{align}
where $f(r) = g(r) = \left(1 - \frac{M}{r}\right)^2$, and the scalar field solution is $\Phi(r) = \frac{\sqrt{3}M}{r - M}$. The horizon is located at $r_h=M$, and it is obvious that the scalar field is divergent at the horizon. 

In fact, this black hole metric coincides with that of an extremal charged black hole in Einstein-Maxwell theory with $Q = M$. For such cases with $\beta \equiv 1/T \to \infty$, it is inappropriate to define the Euclidean integrals. However, this model serves as a good opportunity to study how the divergent behavior of the scalar field at the horizon affects the equality \eqref{xiaoformula}. Thus, we temporarily omit the integration over the time periodicity, which is irrelevant in essence, and only consider the spatial part of the integral $I/\beta \equiv \int_{\mathcal{V}} d^3x\sqrt{|g|}L$.

Then we examine the behavior of the integral
\begin{align}
\int_{S_{r}} n_\mu K^{\mu} \sim \int_{S_{r}} g^{rr}n_{r}\nabla_r (\Phi^2) \sim \frac{M^2}{M - r}.
\end{align}
While the integral still vanishes as $r \to \infty$, it does not vanish as $r \to r_h$ because the derivative of the scalar fields diverges faster than $g^{rr}$ approaches zero. Explicit calculation shows
\begin{align}
I_{\text{tot}}/\beta \sim  \int_{\mathcal{V}_\epsilon} L_{\text{tot}} \sim \frac{M^2}{\epsilon}\sim \infty,
\end{align}
where the integral region $\mathcal{V}_{\epsilon}$ spans from $r = r_h + \epsilon$ to infinity. Thus, we have $(2 \cdot 1 - 4)\, I_{\text{tot}}/\beta \neq 0$. This nonzero outcome signals the divergent behavior of the scalar field at the horizon.

\section{Concluding remarks} \label{sec5}

In this paper, we have investigated the equality relation among Euclidean integrals given by eq.\eqref{xiaoformula}, provided its proof, and generalized it to asymptotically AdS spacetimes as well as scenarios involving couplings between matter fields and gravity. What is particularly interesting is that, despite the wide variety of effective Lagrangians encountered in the literature and the often highly complex nature of the calculated Euclidean integrals, a simple and universal relationship exists among these intricate results.

The significance of the formula \eqref{xiaoformula} and its AdS generalization \eqref{xiaoforads} lies in revealing a fundamental constraint among Euclidean integrals of different action terms, showing that they are not independent of each other. By rearranging the formula to isolate the integral of the Ricci scalar, we obtain
\begin{align}
\alpha_1 I_1 = - \sum_{m \neq 1} \frac{2m - D}{2 - D} \alpha_m I_m,
\end{align}
or more explicitly:
\begin{align}
\int_{\mathcal{M}} \frac{R}{16\pi} = \sum_{m \neq 1} \frac{2m - D}{D - 2} \int_{\mathcal{M}} \alpha_m L_m.
\end{align}
In analyzing black hole thermodynamics, one often needs to calculate the total Euclidean integral. According to the above relation, we can show that
\begin{align}
\begin{split}
\int_{\mathcal{M}} L_{\text{tot}} &= \int_{\mathcal{M}} \frac{R}{16\pi} + \sum_{m \neq 1} \int_{\mathcal{M}} \alpha_m L_m \\
&= \sum_{m \neq 1} \frac{2(m - 1)}{D - 2} \int_{\mathcal{M}} \alpha_m L_m. \label{ltotlm}
\end{split}
\end{align}
The above formula has deep physical implications when analyzing the fundamental laws of extended black hole thermodynamics \cite{Xiao:2023lap}. As is known, the Smarr relation is important in that it connects different thermodynamic quantities of a black hole. For a static and spherical black hole in higher derivative gravity theories, we can derive \( M_K - 2TS = -2 \int_{\mathcal{M}} L_{\text{tot}} \), where \( M_K \) denotes the generalized Komar mass. The formula \eqref{ltotlm} explicitly shows that the total Euclidean integral is proportional to the coupling parameters of higher-order terms (i.e., \( \alpha_m \) with \( m \neq 1 \)). In this way, it plays a central role in constructing the Smarr relation for higher derivative gravity theories; see Ref.\cite{Xiao:2023lap} for details. A further remark is that the Euclidean action encodes information about the system, and our relation shows that this information appears to be distributed among the various terms of the action according to a specific proportional relationship.

Our general analysis indicates that the formulas \eqref{xiaoformula} and \eqref{xiaoforads} hold universally in pure gravity, applying to both asymptotically flat and AdS cases. For the interested reader, Appendix B provides a further analysis of their generalization to de Sitter black holes. When matter fields are coupled to gravity, the equality persists if the matter fields behave properly, i.e., they are non-divergent at the horizon and decay appropriately at infinity. The Maxwell field serves as an example of such well-behaved matter, where the equality is explicitly verified.

Furthermore, we have studied counter-examples to eqs.\eqref{xiaoformula} and \eqref{xiaoforads} where matter fields exhibit pathological behaviors, i.e., divergence at the horizon or failure to decay to zero at infinity. In fact, when such behaviors occur, the validity of relating the Euclidean action to black hole thermodynamics often becomes questionable. For instance, defining the Euclidean action becomes problematic when scalar fields diverge at the horizon. For black holes with a Kalb-Ramond field background, naively using the Euclidean action as a partition function does not yield thermodynamic quantities consistent with those from Wald method \footnote{The ambiguity of the Euclidean action can be illustrated as follows. According to eq.\eqref{krnk}, we have $ \int_{\mathcal{M}} \nabla_\mu K^{\mu}\sim \int_{\mathcal{M}} \nabla^{\nu}(B^{\mu\rho}B_{\nu\rho}) \sim b^2 \xi_2 m$. We can then add such a total-derivative term to the Lagrangian. This procedure does not alter the black hole solution but changes the calculated results of the Euclidean action. Consequently, we cannot reasonably relate the ambiguous Euclidean action to the thermodynamic partition function.}, similar to observations in Einstein--Horndeski theories \cite{Feng:2015oea,Feng:2015wvb}. Thus, the failure of the formulas \eqref{xiaoformula} and \eqref{xiaoforads} can signal anomalous matter field configurations, implying the breakdown of Euclidean methods, or alternatively, indicating the need for more careful and subtle refinements to properly apply Euclidean path integrals in analyzing black hole thermodynamics.

\section*{Acknowledgments}
 XY would like to thank Yu Tian and Hongbao Zhang for useful discussions. YX is also thankful to the Higgs Centre for Theoretical Physics at the University of Edinburgh for providing research facilities and hospitality during the visit. This work was supported in part by China Scholarship Council with Grant No. 202408130101, the Hebei Natural Science Foundation with Grant No. A2024201012, and the National Natural Science Foundation of China with Grant No. 12475048.

 \section*{Appendix A}
The appendix provides more details about the example \eqref{6dl3} mentioned in the main text. Specifically, we study the Lagrangian $L_{\text{tot}}=\alpha_1 L_1+\alpha_3L_3=\frac{R}{16\pi}+ \alpha_3 R_{\mu\nu\rho\sigma}R^{\rho\sigma\alpha\beta}R_{\alpha\beta}^{\ \ \mu\nu}$ in $D=6$ dimensions, where $\alpha_m=0 $ ($m\neq 1,3$ ).  Treating $\alpha_3 L_3$ as a perturbative correction to Einstein gravity, we derive the black hole metric up to first order in $\alpha_3$, which has the form
\begin{align}
ds^2 = -f(r)dt^2 + \frac{1}{g(r)}dr^2 + r^2 d\Omega_4, \tag{A.1}
\end{align}
with
\begin{align}
\begin{split}
f(r) &= 1 - \frac{m}{r^3} - \frac{\alpha_3 240 \pi m^2}{r^{13}} \left(5m - 24 r^3\right), 
\end{split} \tag{A.2}
\end{align}
and
\begin{align}
\begin{split}
g(r) &= 1 - \frac{m}{r^3} - \frac{\alpha_3 240 \pi m^2}{r^{13}} \left(71m - 90 r^3\right).
\end{split}\tag{A.3}
\end{align}
Here, $m$ denotes the mass parameter. According to the condition $f(r_h) = g(r_h) = 0$, the relation between $m$ and the horizon radius $r_h$ is obtained as
$m = r_h^3 + \frac{4560 \pi \alpha_3}{r_h}$.

Using the metric (A.1), we calculate the Ricci scalar, which is non-vanishing and given by
\begin{align}
R = -\frac{\alpha_3 28800 \pi}{r^{15}} \left(10 m^3 - 7 m^2 r^3\right).\tag{A.4}
\end{align}
Despite the nonzero Ricci scalar, explicit integration over the manifold shows that 
\begin{align}
    \int_{\mathcal{M}} R \sim \frac{\alpha_3 m^2 \left(m - r_h^3\right)}{r_h^{10}}.\tag{A.5}
\end{align} Substituting the relation between $m$ and $r_h$ reveals that $\int_{\mathcal{M}} R = 0$ at $\mathcal{O}(\alpha_3)$. Besides, we have further verified this result by solving the metric perturbatively to higher orders in $\alpha_3$, confirming that $\int_{\mathcal{M}} R = 0$ holds, even though  $R \neq 0$ in general. This is consistent with our formula \eqref{xiaoformula}.

\section*{Appendix B}

This appendix provides a concise analysis of the generalization of eqs.\eqref{xiaoformula} and \eqref{xiaoforads} to the Euclidean integrals for de Sitter black holes. 

Actually, there have long been many subtleties in the study of the thermodynamics of de Sitter black holes \cite{Morvan:2022ybp,Banihashemi:2022htw}. Unlike the asymptotically flat and asymptotically AdS cases, de Sitter black holes lack a suitable asymptotic region to facilitate the definition of various conserved quantities. Furthermore, for a general de Sitter black hole, the black hole horizon and cosmological horizon possess different temperatures, so the entire spacetime is often regarded as being in a non-equilibrium state. A consequent result is that we cannot eliminate the conical singularities at both horizons simultaneously. However, Ref. \cite{Morvan:2022ybp} points out that we can still calculate the Euclidean action of the entire spacetime in static coordinates; when combined with contributions from the conical singularities, the total Euclidean action is exactly equal to minus the sum of the entropies of the black hole horizon and the cosmological horizon. Then the total Euclidean action can be interpreted as the partition function of a microcanonical system with fixed mass (independent of temperature). For a detailed physical analysis, refer to \cite{Morvan:2022ybp}.

We focus primarily on the relationship among the Euclidean integrals of different action terms. For the simplest Einstein gravity, the Euclidean integral contains only two terms: $\alpha_0 L_0 = -\frac{\Lambda}{8\pi}$ ($\Lambda=\frac{3}{l^2}>0$) and $\alpha_1 L_1 = \frac{R}{16\pi}$. The trace equation contains no total derivative terms, thus integrating the trace equation directly leads to $\sum_{m=0,1} (2m-D) \alpha_m I_m = 0$, which is a rather trivial situation. To better explore this pattern, we still need to consider the problem within the framework of higher derivative gravity. When the action includes higher derivative terms, the trace equation (see eq.\eqref{traceeq}) will contain a total derivative term of the form $\nabla_\mu K^\mu$. The integral of the total derivative term over the entire spacetime transforms into fluxes at the black hole horizon and cosmological horizon, which vanish due to the property $g^{rr}=0$ of the horizons. Therefore, we naturally expect that the formulas \eqref{xiaoformula} and \eqref{xiaoforads} still hold when generalized to de Sitter spacetimes.

To make the above analysis more concrete, we consider an example described by the Lagrangian: 
\begin{align}
L_{\text{tot}} = \frac{1}{16\pi}\left(R - 2\Lambda\right) + \alpha_3 R_{\mu\nu}^{\ \ \rho\sigma}R_{\rho\sigma}^{\ \ \alpha\beta}R_{\alpha\beta}^{\ \ \mu\nu}. \tag{B.1}
\end{align}
Treating the last term $\alpha_3 L_3$ as a perturbation, we can solve the gravitational equations around the original de Sitter black hole. Up to the first order in $\alpha_3$, the solution in static coordinates is given by: 
\begin{align}
ds^2 = -f(r)dt^2 + \frac{1}{g(r)}dr^2 + r^2 d\Omega_2, \tag{B.2}
\end{align}
where $f(r) = 1 - \frac{2M}{r} - \frac{r^2}{l^2} + \alpha_3\, f^{(1)}(r)$ and $g(r) = 1 - \frac{2M}{r} - \frac{r^2}{l^2} +  \alpha_3 \, g^{(1)}(r)$, with
\begin{align}
\begin{adjustbox}{width=\linewidth-3.28em}
$ \displaystyle
f^{(1)}(r)  \!  = \! \frac{64 \pi }{l^6 r^7} \left(10 l^6 M^3 \!- \! 12 l^4 M^2 r^3  \!- \! 6 l^2 M r^6  \!- \! r^9\right),
$
\end{adjustbox}\tag{B.3}
\end{align}
\begin{align}
\begin{split}
g^{(1)}(r) &=  \frac{64 \pi }{l^6 r^7} \left(54 l^6 M^2 r - 98 l^6 M^3 - 66 l^4 M^2 r^3 \right. \\
&\quad \left. - 6 l^2 M r^6 - r^9\right).
\end{split} \tag{B.4}
\end{align}
When the mass satisfies $0 < M < M_N$, this spacetime has two horizons: the black hole horizon at $r = r_h$ and the cosmological horizon at $r = r_c$, where $M_N$ denotes the Nariai limit at which $r_h$ and $r_c$ coincide. We now consider the Euclidean integral $I_m = \int_{\mathcal{M}} L_m$ integrated from $r_h$ to $r_c$, and examine the specific combination: $(2 \cdot 0 - 4)\alpha_0 I_0 + (2 \cdot 1 - 4)\alpha_1 I_1 + (2 \cdot 3 - 4)\alpha_3 I_3$. Up to the first order in $\alpha_3$, the calculation is straightforward, and the outcome is 
\begin{align}
\left. \alpha_3 \frac{6912 \pi M^2}{r^6}  \left( M - \frac{r}{2} + \frac{r^3}{2 l^2} \right) \right|_{r_h}^{r_c}. \tag{B.5}
\end{align}
Due to the relations between the mass and the two horizon radii: $M = \frac{r_h}{2} - \frac{r_h^3}{2 l^2} + \mathcal{O}(\alpha_3)$ and $M = \frac{r_c}{2} - \frac{r_c^3}{2 l^2} + \mathcal{O}(\alpha_3)$, the final outcome vanishes at $\mathcal{O}(\alpha_3)$.

Note that in some studies, the black hole horizon and cosmological horizon are separated by an arbitrarily chosen timelike boundary at $r = r_y$, such that the black hole horizon can be considered to be in a cavity that maintains thermal equilibrium with it (see e.g., \cite{Banihashemi:2022htw}). Thus the Euclidean action involves an integral from $r_h$ to $r_y$. In this case, the corresponding integral produces 
\begin{align}
\left. \alpha_3 \frac{6912 \pi  M^2}{r^6}   \left( M - \frac{r}{2} + \frac{r^3}{2 l^2} \right) \right|_{r_h}^{r_y}. \tag{B.6}
\end{align}
It is evident that substituting $r = r_y$ does not make the result zero, unless $ r_y $ lies at one of the horizons. This reflects the main argument of this paper: for a well-defined Euclidean integral, the combination $\sum_m (2m-D) \alpha_m I_m = 0$ holds. When a non-zero result emerges, it often indicates some form of unnaturalness in the system; here, this unnaturalness arises from the artificially arbitrary choice of the boundary surface.

\end{document}